# Configurable Electrostatically Doped High Performance Bilayer Graphene Tunnel FET

Fan W. Chen*, Hesameddin Ilatikhameneh*, Gerhard Klimeck, Zhihong Chen, Rajib Rahman

**Abstract**— A bilayer graphene based electrostatically doped tunnel field-effect transistor (BED-TFET) is proposed. Unlike graphene nanoribbon TFETs in which the edge states deteriorate the OFF-state performance, BED-TFETs operate based on bandgaps induced by vertical electric fields in the source, channel, and drain regions without any chemical doping. The performance of the transistor is evaluated by self-consistent quantum transport simulations. This device has several advantages: 1) ultra-low power ($V_{DD}$=0.1V), 2) high performance ($I_{ON}/I_{OFF}>10^4$), 3) steep subthreshold swing (SS<10mv/dec), and 4) electrically configurable between N-TFET and P-TFET post fabrication. The operation principle of the BED-TFET and its performance sensitivity to the device design parameters are presented.

## I. INTRODUCTION

It has been experimentally challenging to realize a tunnel FET (TFET) with high on-current and a steep subthreshold slope simultaneously, especially with a low supply voltage ($V_{DD}$~0.1V). The high current can be achieved by bringing the transmission probability through the source-channel tunneling barrier close to unity, which can be realized by minimizing the effective mass of the channel material and the screening length [1, 2] across the tunneling barrier. Regarding the requirement of small effective mass, bilayer graphene (BLG) is almost an ideal candidate. However, despite its small effective mass, impressive mobility and initial promise for high performance electronic devices [3, 4], the lack of an intrinsic band gap prevents graphene transistors from switching off. Although sizable bandgaps were demonstrated in graphene nano-ribbons (GNRs) [5-8], the edge roughness and device-to-device variations due to the lack of atomic level control in top down fabrication pose a tremendous challenge for technology development [7, 9-11]. On the other hand, a tunable bandgap larger than 200meV can be created in BLG by an electric field [12-14].

Here, BED-TFET as a high performance steep SS device which enables $V_{DD}$ to scale down below 0.1V is proposed. Accordingly, an excellent energy-delay product is obtained in this device. Compared to previous bilayer graphene TFET designs [13, 15], BED-TFET has the following advantages: 1) Being electrostatically configurable post fabrication between a P-TFET and a N-TFET. 2) Avoiding the experimentally challenging chemical doping in 2D materials (i.e. bilayer graphene). 3) Being immune to threshold variations due to dopant fluctuations which is critical for low threshold voltages. 4) Avoiding dopant states within the bandgap which deteriorates the OFF-state performance of the TFETs [16]. 5) Providing an artificial heterostructure without interface states.

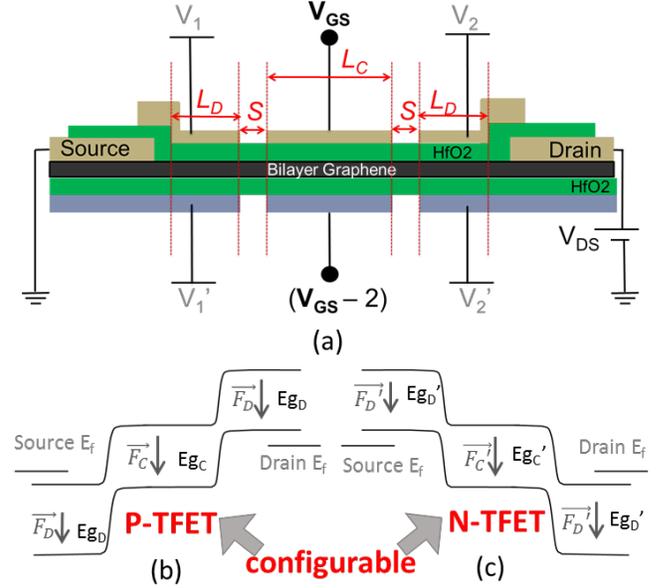

Fig. 1: a) Physical structure of an electrically doped p-i-n BLG TFET. The band diagram in the OFF state of BED-TFET as a b) N-TFET, c) P-TFET.

The device structure is shown in Fig. 1(a). The left and right regions are controlled by $V_1$, $V_1'$ and $V_2$, $V_2'$, respectively, and act as the electrostatically doped source and drain regions for the TFET. By adjusting $V_1...V_2'$, the proposed device is configurable between an N-TFET and a P-TFET as shown in Fig. 1(b) and 1(c). The bandgap size of each region is also tunable by the voltage difference ($\Delta V$) between the top and the bottom gates in that region. The induced band gaps are denoted by $Eg_C$ and $Eg_D$. Accordingly, an artificial heterostructure can be made as long as the electric fields of different regions are different, $\vec{F_D} \neq \vec{F_C}$. The fabrication of BED-TFET requires the

* These two authors contributed equally.
This work was supported in part by the Center for Low Energy Systems Technology (LEAST), one of six centers of STARnet, a Semiconductor Research Corporation program sponsored by MARCO and DARPA.

The authors are with the Network for Computational Nanotechnology (NCN), Purdue University, West Lafayette, IN, 47906 USA e-mail: fanchen@purdue.edu, hesam.ilati2@gmail.com

alignment of top and bottom gates, which can be challenging. However, advanced workfunction engineering techniques[17] may be used to reduce the number of gates, however, a detailed investigation of such technique are beyond the scope of this paper.

One of the main advantages of the BED-TFET is its very low energy-delay product. Fig. 2 benchmarks the energy-delay of a 32 bit adder [18] based on different steep devices. The benchmarking methodology is described in [18] for beyond-CMOS devices. The BED-TFET has the least energy-delay product among the studied devices. This is due to the steep IV and high $I_{ON}$ obtained in the BED-TFET even with a low $V_{DD}$ of 0.1V. This shows the importance of low band gap materials for low $V_{DD}$ steep devices. Notice that the parasitic capacitances between the gates can be significantly reduced by using a low-k dielectric ($\epsilon_S$) between the gates [1] and increasing the spacing ($S$); e.g. a 10nm air gap spacer can reduce parasitic capacitances about 2 orders of magnitude smaller than gate capacitance ($\epsilon_S/S \ll \epsilon_{ox}/t_{ox}$). According to Fig. 2, this parasitic capacitance doesn't degrade the energy-delay product of BED-TFET.

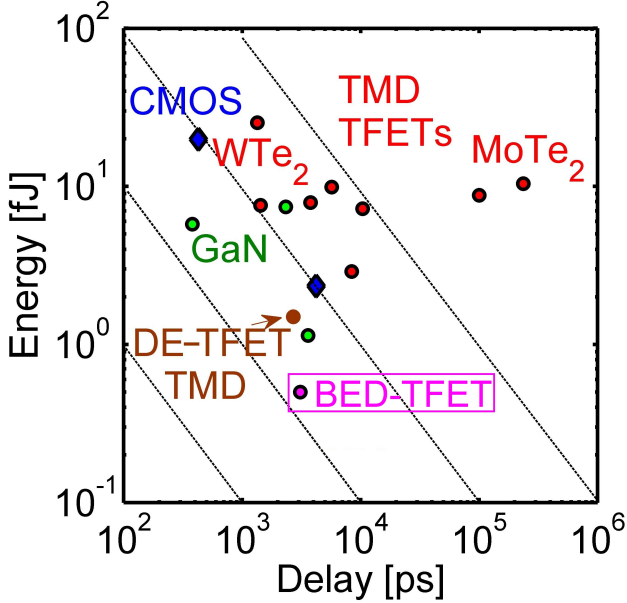

Fig. 2: Energy-Delay comparison of BED-TFET (pink dot) with Dielectric Engineered (DE) WTe2 TFET (brown dot) [1], Nitride TFET (green dots) [19], TMD TFETs (red dots) [2, 20, 21] and Si MOSFET (blue dots) [18, 22].

## II. SIMULATION DETAILS

The Hamiltonian of BLG is represented using a $p_z$ orbital nearest-neighbor tight-binding (TB) model, which contains only in-plane and inter-plane hopping terms, $\gamma_0$ and $\gamma_1$ as listed in Table I. The material properties of the BLG under vertical field extracted from the bandstructure for the maximum $Eg$ of 275 meV are also in Table I. All the transport characteristics of the BED-TFET have been simulated using the self-consistent Poisson-Non Equilibrium Green's Function (NEGF) method through the Nano-Electronic MOdeling (NEMO5) tool [23-31]. Applying a vertical field to BLG opens up a band gap (Fig. 3).

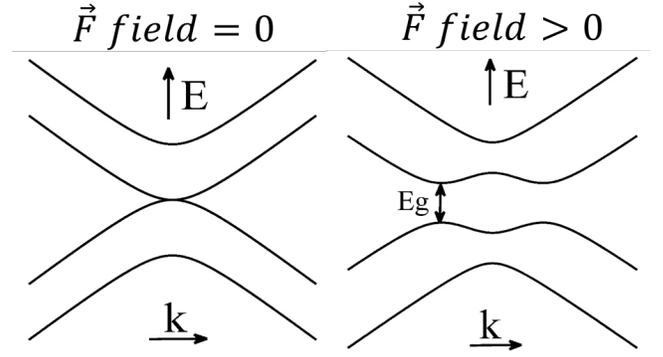

Fig. 3: Vertical electric field opens up a bandgap in BLG.

The BED-TFET shown in Fig. 1a) is composed of a bilayer graphene layer sandwiched between two layers of 3nm thick HfO$_2$ with a relative dielectric constant of $\varepsilon_r = 20$. The maximum field within HfO$_2$ in current BED-TFET design is about 3MV/cm which is less than the breakdown field of HfO$_2$ (~8.5MV/cm) [32]. The three gated regions from left to right have lengths of 25, 40 and 25 nm. $\Delta V$ in the middle region is fixed to 2V to reach the maximum bandgap (i.e. 275meV in BLG). Notice that, $V_1 \ldots V_2'$ are fixed throughout the device operation to achieve the desired electrostatically doping. Only the gate voltages in the middle region are swept to switch the device between ON and OFF.

Table I: Bilayer graphene material properties: in-plane and inter-plane hopping parameters $\gamma_0$ and $\gamma_1$, maximum bandgap $Eg$, electron effective mass $m_e^*$, in-plane and out-plane relative dielectric constant $\epsilon_r^{in}$ and $\epsilon_r^{out}$.

| Parameters | $\gamma_0$ (eV) | $\gamma_1$ (eV) | $Eg$ (meV) | $\epsilon_r^{in}$ | $\epsilon_r^{out}$ |
|---|---|---|---|---|---|
| Bilayer Graphene | 2.75 | 0.3 | 275 | 3 | 3.3 |

## III. RESULTS AND DISCUSSION

All the results here are for BED-TFET with P-FET configuration in Fig. 1(b); $V_1$, $V_1'$ and $V_2$, $V_2'$ are fixed at 1.1V, -0.1V and 0.4V, -0.8V respectively to form the electrostatically doped source and drain regions.

Fig. 4(a) shows the local band diagram along the transport direction (left) and energy resolved current for the ON-state (right) of the device. There is a tunnel window of about 210 meV in the ON-state. Due to the small band gap at the tunnel junction, the ON-current is high. In the OFF-state, the middle region blocks the tunneling window as shown in Fig. 4(b). Consequently, the OFF-current is mainly the result of the thermionic electron and hole currents. The electrically induced band gaps at the source and drain regions in conjunction with the band gap of the channel make an effective barrier height of about 350meV which is large enough to reduce the thermal current at 300K to the desired range.

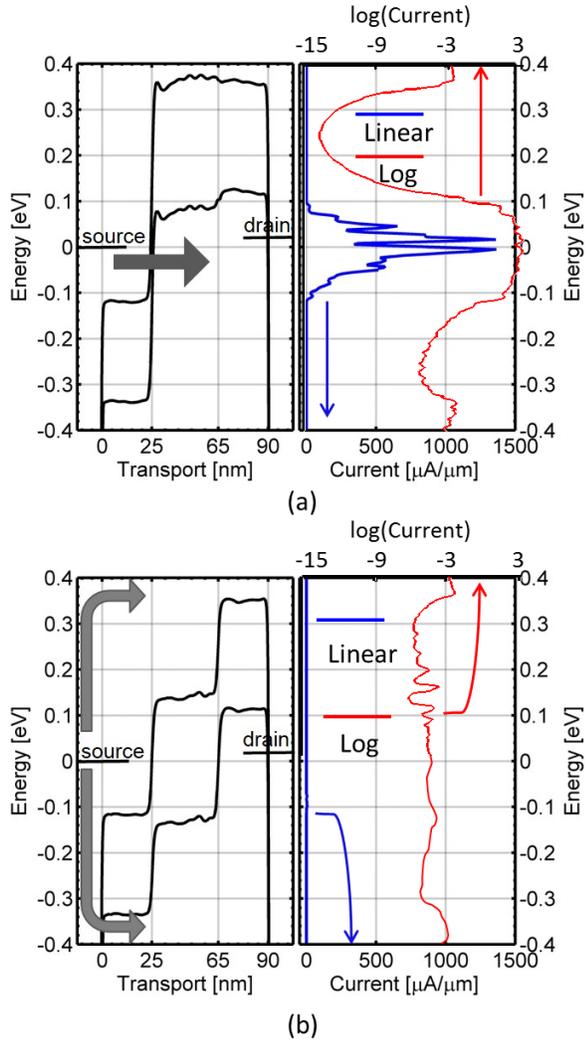

Fig. 4: The band diagram along the transport direction (left) and the energy resolved current (right) in (a) ON state and (b) OFF state.

Fig. 5(a) shows the transfer characteristics of the BED-TFET for different $V_{DS}$ values. Increasing $|V_{DS}|$ from 10mV to 100mV increases both the ON and OFF currents. Fig. 5(b) shows that this device achieves a small SS value of 8 mV/dec and high $I_{60}$ (the current value where SS becomes 60 mV/dec) value of 24 µA/µm for a $V_{DS}$ of -100mV. Notice that this value of $I_{60}$ is much higher than that of other 2D material TFETs even with a $V_{DD}$ of 0.5V [21]. Notice that increasing $|V_{DS}|$ from 10mV to 100mV does not deteriorate the small SS. Fig. 5(c) plots the output characteristics of the device. $I_D$-$V_{DS}$ curves show that there is no late turn on problem in BED-TFET and the linear region of $I_D$-$V_{DS}$ starts from $V_{DS}$=0V. Moreover, the current saturates for $|V_{DS}|$ values above 50mV. Fig. 5(d) shows that an increase in $|V_{DS}|$ decreases the ON/OFF ratio from $5\times10^4$ at $V_{DS}$=-10mV to $2\times10^4$ at $V_{DS}$=-100mV, which is not substantial.

Here, tunnel thickness modulation rather than energy filtering is used to achieve steep slope. However, unlike other TFETs that operate with tunnel thickness modulation, the bandgap is dictated locally by the vertical field which is smaller at the source-channel interface than in the channel. Consequently, a larger current can be achieved in this TFET.

Notice that the energy filtering mechanism is not effective in low bandgap materials since the small gap can only block a small portion of the Fermi tail.

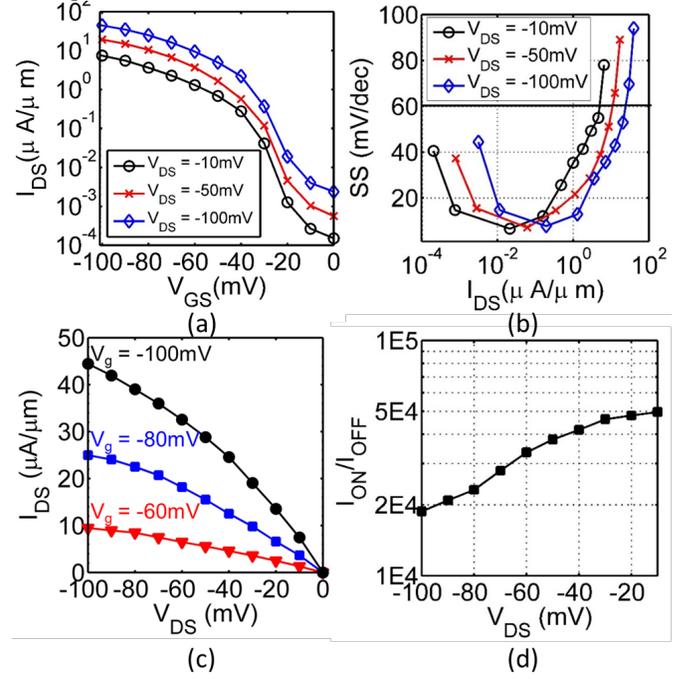

Fig. 5: a) Transfer characteristics of the BED-TFET with different drain-to-source voltages $V_{DS}$. b) SS-Id plot with different drain-to-source voltages $V_{DS}$. c) Output characteristics of the TFET at several gate voltages $V_g$. d) ON/OFF ratio with source-drain voltage $V_{DS}$ for the BED-TFET. $L_C$, $L_D$ and $S$ are kept at 40nm, 25nm and 0nm, respectively.

In the BED-TFET, several design parameters are identified to be critical for the device performance and fabrication: 1) the channel length $L_C$, 2) the length of the electrostatically doped source and drain regions $L_D$, and 3) the spacing between these gated regions $S$. In the transfer characteristics demonstrated in Figs. 6a-c, $L_C$, $L_D$ and $S$ are kept at 40nm, 25nm and 0nm respectively, unless mentioned otherwise. Fig. 6(a) shows that reducing $L_C$ to 40nm increases the OFF-current. Below, the performance is not sensitive to S as shown in Fig. 6(b) for S in the range of 0nm to 20nm. Fig. 6(c) shows that a $L_D$ value below 25nm can impact the OFF-state performance. The sensitivity to $L_C$ and $L_D$ originates from the direct tunneling of carriers through the channel potential barrier due to the small effective mass of the BLG. The optimized channel length is longer than the ITRS requirements. Hence, to keep the footprint of the BED-TFET small a vertical structure (e.g. conventional vertical TFET structure [33]) could be used.

## IV. Conclusion

In this work, the BED-TFET is proposed as a high performance, ultra-low power, steep transistor to overcome the problems associated with GNRs. The electrically tunable band gap of BLG makes this transistor highly configurable. The performance of this device is evaluated through rigorous quantum transport simulations based on NEGF. It is shown that with the right device design, the BED-TFET can achieve ON/OFF ratios of more than $10^4$, ON-current of 45µA/µm, and a subthreshold swing around 10 mV/dec, all at a low overdrive voltage of $V_{DD}$ =0.1V at room temperature.

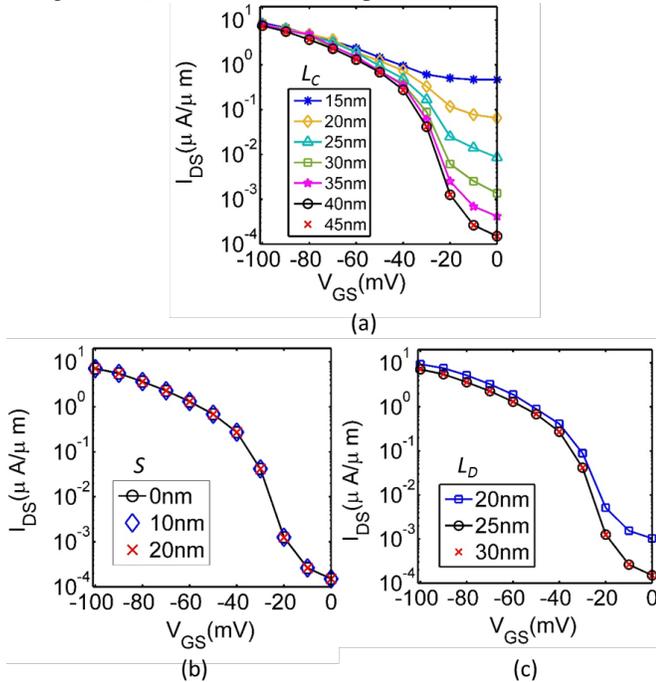

Fig. 6: $L_D$, $L_C$ and $L_D$ are the gate length of the left/middle/right region in Fig. 1(a) (doping region/channel/doping region), respectively. The spacing between the gates is $S$. Transfer characteristics of the TFET with different a) channel length $L_C$ ($L_D$ = 25nm, $S$=0nm), b) spacing $S$ ($L_C$ = 40nm, $L_D$ =25nm) and c) doping region length $L_D$ ($L_C$ = 40nm, $S$ = 0nm).

ACKNOWLEDGMENT

The authors would like to thank J. Nahas and R. Perricone for the 32-bit adder energy-delay calculations.